\documentclass[pra,preprint,aps]{revtex4}
\begin{document}

\title{On the Quintessence Scalar Field Potential}

\author{J. A. Espich\'an Carrillo} \email{jespichan@unac.edu.pe}
\affiliation{Facultad de Ciencias Naturales y Matem\'atica,
Universidad Nacional del Callao (UNAC) - Bellavista - Callao,
Per\'u.}
\author{J. M. Silva} \email{jmsilva@astro.iag.usp.br}
\author{J. A. S. Lima} \email{limajas@astro.iag.usp.br}
\affiliation{Instituto de Astronomia,  Geof\'{\i}sica e Ci\^encias
Atmosf\'ericas, USP, S\~ao Paulo, SP, Brazil}

\date{\today}



\begin{abstract}
\hspace*{0.5cm}In this work we propose a new analytical method for
determining the scalar field potential $V(\phi)$ in FRW type
cosmologies containing a mixture of perfect fluid   plus a
quintessence scalar field. By assuming that the equation of state
parameters of the perfect fluid, $\gamma -1 \equiv
p_{\gamma}/\rho_{\gamma}$ and the  quintessence, $\omega \equiv
p/\rho$ are constants, it is shown that the potential for the flat
case is $V(\phi) = A\rho_{\phi_{0}}\sinh^{B}(\lambda \phi)$, where
$A$, $B$ and $\lambda$ are functions of $\gamma$ and $\omega$. 
This general result is a pure consequence of the Einstein field equations
and the constancy of the parameters. Applying the same method for
closed and open universes, the corresponding scalar field potentials
are also explicitly obtained for a large set of values of the free
parameters $\gamma$ and $\omega$. A formula yielding the transition
redshift from a decelerating to an accelerating regime is also
determined and compared to the $\Lambda$CDM case.
\end{abstract}

\maketitle

\section{Introduction}

\hspace*{0.6cm} A large number of recent observational data strongly
suggest that we live in a flat, accelerating Universe composed of
$\sim$ 1/3 of matter (baryonic + dark) and $\sim$ 2/3 of an exotic
component with large negative pressure, usually named Dark Energy or
Quintessence. The basic set of experiments includes: the luminosity
distance from SNe Ia \cite{Riess}, temperature anisotropies of the
cosmic background radiation \cite{Spergel}, large scale structure,
X-ray data from galaxy clusters \cite{Allen}, age estimates of
globular clusters \cite{Krauss03} and the ages from old high
redshift galaxies \cite{Ages}.

A traditional candidate for the missing energy component is the
vacuum energy density or cosmological constant ($\Lambda$) which is
equivalent to a perfect fluid obeying the equation of state $p_v = -
\rho_v$. Due to the cosmological constant problem \cite{Wein}, some
authors have also considered that the vacuum energy density, due to
its coupling with the other matter fields, can be a time-dependent
function ($\Lambda(t)-models$) \cite{Lambda}.

A more generic possibility is a dynamical, time-dependent scalar
field $\phi$ evolving slowly  in its potential $V(\phi)$, which is
usually referred to as Quintessence field \cite{Quint}. Actually,
due to its simplicity, a scalar field works like  a kind of
paradigma in particle physics (including string theory), an these
can act as dark energy candidates. Although still lacking
experimental evidence of its existence scalar fields are needed in
all unification theories.

The Quintessence cosmological model considered here is defined by
the action
$S={m^2_{pl}/16\pi}\int d^4 x \sqrt{-g}[R
-{1\over2}\partial^{\mu}\phi\partial_{\mu}\phi-V(\phi)+{\cal{L}}_{m}]$,
where $R$ is the Ricci scalar and $m_{pl}\equiv G^{-1/2}$ is the
Planck mass. The scalar field minimally coupled to gravity is
assumed to be homogeneous, such that $\phi=\phi(t)$ and the
Lagrangian density ${\cal{L}}_{m}$ includes the additional perfect
fluid component.

In a cosmological setting it proves convenient to characterize the
scalar field  by an effective equation of state (EoS) parameter,
$w(t)\equiv p/\rho$, measuring the ratio between its pressure and
energy density which is different from baryons, dark matter,
neutrinos  or radiation. Depending on the form of the potential
$V(\phi)$, $w$ can be constant, monotonically increasing
(decreasing) or even oscillatory \cite{R4a}. Actually, a  wide
variety of scalar field models with many disparate applications in
cosmology have been proposed in the literature
\cite{Quint,R4a,caldprl,Applic,R11}.

Whether the EoS parameter $w$ of the Quintessence is constant and
satisfies $\omega\geq -1$, the quintessence has been termed
``X-matter" \cite{R5,R5a}, which also includes  the cosmological
constant ($\Lambda$CDM) models as a limiting case. For these
``X-matter" models the first constraints on the free parameters were
obtained by Perlmutter, Turner and M. White \cite{R7a}. The latest
constraints from  cosmic microwave background anisotropies
\cite{Spergel} alone yields $\Omega_x = 0.73^{+0.10}_{-0.11}$ and
$\omega = -1.06^{+0.41}_{-0.42}$ ($95\%$ C.L.).  The large scale
structure \cite{Allen,R6} provides  $\Omega_m = 0.28 \pm 0.06$ and
$\omega = -1.14 \pm 0.31$ for a flat cosmology while more tight
limits are obtained from SN Ia data \cite{Riess} $\omega = -1.023
\pm 0.090 (stat)\pm 0.054 (sys)$. When $\omega\leq -1$ the
quintessence field is called a phantom fluid \cite{Phantom}. The
basic difficulties and the main advantages underlying the physics of
the different dark energy candidates has been reviewed by several
authors \cite{review}.

In this article we focus our attention to a  Quintessence dark
energy in its X-matter version. The main aim here is to determine
the general analytic form of the scalar field potential which is
simultaneously compatible with the ``X-matter" constraint and the
symmetries of the FRW line element. Actually, this problem has
previously been considered in the literature, however, only special
solutions has been derived \cite{R8,R9,R9a}.

As we shall see, if the ``X-matter" interacts only gravitationally,
that is, in the absence of decaying process or energy transference
among the components, only a very restricted class of potentials is
mathematically allowed by the Einstein Field Equations (EFE). In
this case, the complete spectrum of solutions (for the flat case)
can be fully determined with basis on the  new method proposed here.
In particular, for specific values of the free parameters, the
solutions are slightly different from some expressions recently
obtained in the literature. For closed and hyperbolic Universes,
analytical solutions are also presented for specific values of the
free parameters.

\section{Basic Equations}
We shall restrict our analysis to homogeneous and isotropic
cosmologies described by the FRW line element
\begin{equation}
\label{LE} ds^2 \ = \ dt^2 \ - \ R^2(t)\left[\frac{dr^2}{1-kr^2} +
r^2d\theta^{2} + r^{2}\sin^{2}\theta d\phi^{2}\right]
\end{equation}
where $R(t)$ is the scale factor and $k=0,\pm1$ is the curvature
parameter.

Let us now consider a Universe filled with a perfect fluid plus a
decoupled scalar field $\phi$. In the background (\ref{LE}), the
Einstein's field equations (EFE) can be written as
\begin{equation}
\label{pq2}\frac{8\pi}{m_{pl}^{2}}\,(\rho_{\gamma} + \rho_{\phi}) \
= \ 3\,\frac{{\dot R}^2}{R^2} \ + \  3\,\frac{k}{R^{2}},
\end{equation}
\begin{equation}
\label{pq3}\frac{8\pi}{m_{pl}^{2}}\,(p_{\gamma} + p_{\phi}) \ = \ -
\ 2\,\frac{{\ddot R}}{R} \ - \  \frac{{\dot R}^2}{R^2} \ - \
\frac{k}{R^{2}},
\end{equation}
where an overdot  means time derivative, and $m^{2}_{pl} = 1/G$ is
the Planck mass. The quantities $\rho_{\gamma}$, $\rho_{\phi}$,
$p_{\gamma}$ and $p_{\phi}$ are the energy densities and pressures
of the  perfect fluid and scalar field $\phi$, respectively.

It will be assumed that the perfect fluid and scalar field $\phi(t)$
obeys the following equation of state
\begin{equation}
\label{pq5}p_{\gamma}  =  (\gamma -1)\rho_{\gamma} \ \ \ \ \ \ \
\mbox{and} \ \ \ \ \ \ \ p_{\phi} = w\rho_{\phi}
\end{equation}
where $p_{\phi}=\frac{1}{2}\dot{\phi}^{2}-V(\phi)$ and
$\rho_{\phi}=\frac{1}{2}\dot{\phi}^{2}+V(\phi)$. Moreover, the
constant parameter $\gamma \in [0,2]$ and $w$ take values within the
interval $(-1,0)$ because  the phantom case is not being considered.

On the other hand, since the energy of each component is separately
conserved, the energy densities of both components satisfy
\begin{equation}
\label{pq7}\dot{\rho_{\gamma}} + 3\gamma H \rho_{\gamma} = 0,
\end{equation}
\begin{equation}
\label{pq8}\dot{\rho_{\phi}} \ + \ 3(1 + w) H \rho_{\phi} \ = \ 0,
\end{equation}
where $H =\dot{R}/{R}$ is the Hubble parameter. These equations can
explicitly be integrated giving
\begin{equation}
\label{pq9}\rho_{\gamma} \ = \ \rho_{\gamma
0}\left(\frac{R}{R_{0}}\right)^{-3\gamma} \ \ \ \ \ \ \ \ \
\mbox{and} \ \ \ \ \ \ \ \ \ \rho_{\phi} \ = \
\rho_{\phi_{0}}\left(\frac{R}{R_{0}}\right)^{-3(1 + w)},
\end{equation}
where $\rho_{\gamma 0}$, $\rho_{\phi_{0}}$ and $R_{0}$ are the
values of these parameters at the present time $(t = t_0)$.
Naturally, the second solution is valid only for constant values of
$w$. As usual, inserting the expressions of $\rho_{\phi}$ and
$p_{\phi}$ into the energy conservation law for the scalar field (or
more directly from the field Lagrangian), one obtains the equation
of motion
\begin{equation}
\label{pq4}\ddot{\phi} \ + \ 3H\dot{\phi} \ + \ {dV(\phi) \over
d\phi} = 0.
\end{equation}
If $V(\phi)$ is given a priori, one may follow the standard approach
by integrating directly the above equation. In what follows we
explore the latter approach for the case of ``X-matter".

\section{Evolution of the Scale Factor and the Quintessence Potential}

In order to find the scalar field potential corresponding to a
generic constant EoS parameter in the presence of a perfect fluid,
we start by combining the expressions of $p_{\phi}$ and
$\rho_{\phi}$ defined in (\ref{pq9}). One finds

\begin{equation}
\label{V0}V(\phi) \ = \ \frac{(1-w)}{2(1+w)}\,{\dot{\phi}}^{2} \ \ \
\ \ \ \ \mbox{and} \ \ \ \ \ \ \ \ \rho_{\phi} \ = \
\frac{1}{(1+w)}\,{\dot{\phi}}^2,
\end{equation}
showing that $V(\phi)$ and $\rho_{\phi}$ may be readily determined
if ${\dot{\phi}}^{2}$ is known as a function of $\phi$.

Now, substituting the derivative of $V(\phi)$ with respect to $\phi$
into the equation (\ref{pq4}), one obtains the following
differential equation

\begin{equation}
\frac{\ddot{\phi}}{\dot{\phi}} \ + \
\frac{3(1+w)}{2}\,\frac{\dot{R}}{R} \ = \ 0, \label{pq11}
\end{equation}
and a straightforward integration yields
\begin{equation}
\label{pq13} \dot{\phi} \ = \sqrt{(1+w)\rho_{\phi_{0}}}\,\left(
\frac{R}{R_0}\right)^{-\frac{3(1+w)}{2}} =
\sqrt{(1+w)\rho_{\phi_{0}}}\,x^{-\frac{3(1+w)}{2}}
\end{equation}
where the variable $x=\frac{R}{R_0}$ has been introduced in the
second equality. From (\ref{pq13}), we see that $w= - 1$ implies
$\dot{\phi}=0$. This special case corresponds to the cosmological
constant. The above expression tell us that the solution of our
problem it will be obtained only if the  scale factor is determined
as a function of $\phi$.

On the other hand, by combining the set of equations (\ref{pq2}) -
(\ref{pq5}) and (\ref{pq9}) one finds the differential equation
governing the behavior of the scale factor $R(t)$ in the presence of
a perfect fluid $\gamma$ plus the dark ``X-matter" energy
\begin{eqnarray}
\label{geneq}R{\ddot R} \ + \ \Delta {\dot R}^2 \ + \ \Delta\,k \ +
\ \frac{3}{2}\,H_{0}^{2}(1 - \gamma +
w)\Omega_{\phi_{0}}R_{0}^{3(1+w)} R^{-(1+3w)} \ = \ 0,
\end{eqnarray}
whose first integral can be written as
\begin{eqnarray}
{\dot{R}}^{2}  =  \frac{A}{R^{3\gamma - 2}} \ - \ k \ + \
H_{0}^{2}\Omega_{\phi_{0}}R^{3(1+w)}_{0}R^{-(1+3w)},
\end{eqnarray}
where $A = H_{0}^{2}\,\Omega_{\phi_{0}}\,R^{3\gamma}_{0}$ is an
integration  positive constant e  $\Delta \equiv \frac{3\gamma -
2}{2}$. In the absence of a quintessence component
($\Omega_{\phi}\equiv 0$) the above Eq. (\ref{geneq}) reduces to the
general FRW differential equation as discussed by Assad and Lima
\cite{Lima88}.

On the other hand, from the first EFE (\ref{pq2}) one find that $k$
parameter satisfies $\Omega_{\gamma 0} + \Omega_{\phi_{0}} - 1 =
\frac{k}{H_{0}^{2}R^{2}_{0}}$. As one may check, inserting the
values of $A$ and $k$ into the above equation, and by introducing
the variable $x = R/R_o$ it follows that
\begin{equation}
\label{pq14}{dt \over dx}\ =
\frac{H_{0}^{-1}}{\sqrt{1-\Omega_{\gamma 0}-\Omega_{\phi_0}+
\Omega_{\gamma 0}\, x^{-(3\gamma
-2)}+\Omega_{\phi_0}\,x^{-(1+3w)}}},
\end{equation}
where $H_0$ is the Hubble parameter at the current time $(t = t_0)$.
Thus, substituting (\ref{pq14}) into (\ref{pq13}) we obtain
\begin{equation}
\label{pq15}d\phi \ =  H_{0}^{-1}\,\sqrt{(1+w)\rho_{\phi_0}}\,
\frac{x^{-\frac{3}{2}(1+w)}\,dx}{\sqrt{1-\Omega_{\gamma
0}-\Omega_{\phi_0}+ \Omega_{\gamma 0}\,x^{-(3\gamma
-2)}+\Omega_{\phi_0}\,x^{-(1+3w)}}}.
\end{equation}

The integration and inversion of the above equation yields $R(\phi)$
and from Eq. (\ref{V0}) one obtains the scalar field potential.
However, it cannot be analytically solved for arbitrary values of
the curvature parameter. As discussed below, a general solution
exist only for the flat case. Solutions for $k = \pm 1$ are also
possible for specific values of the pair ($\gamma, \omega$).

\subsection{Solution of the flat case}

For $k=0$ we see that $\Omega_{\gamma 0} + \Omega_{\phi_0} = 1$.
Inserting this into (\ref{pq15}) and introducing the auxiliary
coordinate $\theta$ $\frac{\Omega_{\phi_{0}}}{\Omega_{\gamma
0}}\,x^{3(\gamma-w-1)} \ = \ \sinh^{2}\theta$, one finds
\begin{equation}
\label{pq25}R(\phi) \ = \ R_{0}\,\left( \frac{\Omega_{\gamma
0}}{\Omega_{\phi_{0}}} \right)^{\frac{1}{3(\gamma
-w-1)}}\,\sinh^{\frac{2}{3(\gamma -w-1)}} \left[ \frac{3\,(\gamma -
w -1)\sqrt{8\,\pi}} {2\,\sqrt{3(1+w)}}\,\frac{\phi}{m_{pl}} \right],
\end{equation}
or equivalently,
\begin{equation}
\label{pq26}\frac{\phi(R)}{m_{pl}} \ = \
\frac{2\,\sqrt{3(1+w)}}{3\,(\gamma -w-1)\,\sqrt{8\,\pi}}\, {\rm
arcsinh}\left[ \sqrt{\frac{\Omega_{\phi_{0}}}{\Omega_{\gamma 0}}}
\left( \frac{R}{R_0}\right)^{\frac{3(\gamma -w-1)}{2}} \right],
\end{equation}
where the integration constant has been fixed equal to zero. Now,
inserting (\ref{pq25}) into (\ref{pq13}) and using (\ref{pq4}), we
obtain the following expression for the scalar field potential

\begin{equation}
\label{SG}V(\phi) = \frac{(1-w)}{2}\rho_{\phi_{0}} \left(
\frac{\Omega_{\phi_{0}}}{\Omega_{\gamma
0}}\right)^{\frac{(1+w)}{(\gamma
-w-1)}}\sinh^{-\frac{2(1+w)}{(\gamma -w-1)}} \left[ \frac{3\,(\gamma
- w -1)\sqrt{8\,\pi}} {2\,\sqrt{3(1+w)}}\,\frac{\phi}{m_{pl}}
\right].
\end{equation}

The corresponding energy densities for the perfect fluid $\gamma$
and the scalar field $\phi$ are given by
\begin{equation}
\label{pq28}\rho_{\gamma}(\phi) \ = \ \rho_{\gamma 0}\,\left(
\frac{\Omega_{\gamma 0}}
{\Omega_{\phi_{0}}}\right)^{\frac{\gamma}{\gamma -w-1}}\,
\sinh^{-\frac{2\,\gamma}{(\gamma -w-1)}}\left[ \frac{3\,(\gamma - w
-1)\sqrt{8\,\pi}} {2\,\sqrt{3(1+w)}}\,\frac{\phi}{m_{pl}} \right],
\end{equation}
\begin{equation}
\label{pq29}\rho_{\phi}(\phi) = \rho_{\phi_{0}}\left(
\frac{\Omega_{\phi_{0}}} {\Omega_{\gamma
0}}\right)^{\frac{(1+w)}{(\gamma -w-1)}}\,
\sinh^{-\frac{2(1+w)}{(\gamma -w-1)}} \left[ \frac{3\,(\gamma - w
-1)\sqrt{8\,\pi}} {2\,\sqrt{3(1+w)}}\,\frac{\phi}{m_{pl}} \right].
\end{equation}

Relations (\ref{pq25}) - (\ref{pq29}) are the general and unified
solutions describing the main physical quantities for a flat
universe filled with perfect fluid plus a ``X-matter" component
characterized by the pair ($\gamma, \omega$). Thus, all known
solutions are peculiar cases of it through an adequate choice of the
corresponding parameters. In particular, they allow us to calculate
the expressions at different epochs. For example, substituting
$\gamma = 1$ (dust) and $\gamma = \frac{4}{3}$ (radiation) into
(\ref{SG}), one gets, respectively:
\begin{equation}
\label{pq30}V(\phi) = \frac{(1-w)}{2}\,\rho_{\phi_{0}}\, \left(
\frac{\Omega_{M0}} {\Omega_{\phi_{0}}}\right)^{\frac{(1+w)}
{w}}\,\sinh^{\frac{2(1+w)}{w}} \left[ \frac{- 3\,w\,\sqrt{8\,\pi}}
{2\,\sqrt{3(1+w)}}\,\frac{\phi}{m_{pl}} \right],
\end{equation}

\begin{equation}
\label{pq31}V(\phi) = \frac{(1-w)}{2}\,\rho_{\phi_{0}}\, \left(
\frac{\Omega_{\phi_{0}}}{\Omega_{r 0}}\right)^{\frac{3(1+w)} {(1 -
3w)}}\,\sinh^{-\frac{6(1+w)}{(1 - 3w)}} \left[
\frac{(1-3w)\sqrt{8\,\pi}} {2\,\sqrt{3(1+w)}}\,\frac{\phi}{m_{pl}}
\right].
\end{equation}
Solution (\ref{pq30}) was independently discovered by several
authors using different methods \cite{R9a,R9b,R10}. However, our
general solution (\ref{SG}) display analytically the influence of
the different regimes on the behavior of the potential $V(\phi)$, as
can be seen from the above expression for the radiation phase. More
information may also be obtained taking the limit of (\ref{SG}) at
early times. For ${R}<<{R_0}$ the scalar fields satisfies the
condition $\left| \frac{3(\gamma - w -1)\sqrt{8\pi}}{2\sqrt{3(1+w)}}
\frac{\phi}{m_{pl}}\right| \ll 1$ and from (\ref{SG}) we obtain
\begin{equation}
V(\phi) \sim \frac{(1-w)}{2}\rho_{\phi_{0}}\left(
\frac{\Omega_{\phi_0}} {\Omega_{\gamma 0}}
\right)^{\frac{(1+w)}{(\gamma -w -1)}} \left[ \frac{3(\gamma - w
-1)\sqrt{8\pi}}{2\sqrt{3(1+w)}} \frac{\phi}{m_{pl}}
\right]^{-\frac{2(1+w)}{(\gamma -1-w)}},
\end{equation}
where $\dot{\phi_0}$ was substituted by $\sqrt{(1+w)\rho_{\phi_0}}$.
In particular, for $\gamma = 1$ and $\gamma = \frac{4}{3}$ the
potentials scale as
\begin{equation}
\label{pq32}V(\phi) \sim \left[ \frac{-3w\sqrt{8\,\pi\, \Omega_{M
0}}}{2\,\sqrt{3(1+w)\Omega_{\phi_{0}}}}\, \frac{\phi}{m_{pl}}
\right]^{\frac{2(1+w)}{w}},
\end{equation}

\begin{equation}
\label{pq33}V(\phi) \sim \left[ \frac{(1-3w)\sqrt{8\,\pi\, \Omega_{r
0}}}{2\,\sqrt{3(1+w)\Omega_{\phi_{0}}}}\, \frac{\phi}{m_{pl}}
\right]^{-\frac{6(1+w)}{(1-3w)}},
\end{equation}
which could be obtained directly from equations (\ref{pq30}) and
(\ref{pq31}). As far as we know, the limiting case above for
radiation (\ref{pq33}) was not presented in the literature, whereas
(\ref{pq32}) was first obtained by Ure\~na-L\'opez and Matos
\cite{R10}.

\subsection{Solution $k \neq 0$ (open and closed Universes)}

As remarked before, in this case equation (\ref{pq15}) has no
general analytical solution. Thus we consider arbitrary values of
$\gamma$ with  specific values of $\omega$ and vice-versa. As an
example, let us consider arbitrary $\gamma$ and $w = -\frac{1}{3}$.
In this case, for $w = -\frac{1}{3}$ and introducing the coordinate
$\left[ \frac{1- \Omega_{\gamma 0}} {\Omega_{\gamma
0}}\right]x^{3\gamma-2} = \sinh^{2}\theta$ in (\ref{pq15}), it
reduces to
\begin{equation}
\label{curv} d\phi \ = H_{0}^{-1}\sqrt{\frac{2}{3}\rho_{\phi_0}}\,
\frac{x^{-1}\,dx}{\sqrt{1-\Omega_{\gamma 0}+ \Omega_{\gamma
0}\,x^{-(3\gamma -2)}}}.
\end{equation}
with solution
\begin{equation}
\label{pq18}\frac{\phi(R)}{m_{pl}} \ = \ \frac{1}{(3\gamma
-2)\sqrt{\pi}} \sqrt{\frac{\Omega_{\phi_0}}{1- \Omega_{\gamma 0}}}
{\rm arcsinh}\left[ \sqrt{\frac{1- \Omega_{\gamma 0}}
{\Omega_{\gamma 0}}}\left( \frac{R}{R_0}\right)^{\frac{3\gamma
-2}{2}} \right],
\end{equation}
or equivalently,
\begin{equation}
\label{pq17}R(\phi) \ = R_0 \left(\frac{\Omega_{\gamma 0}}{1-
\Omega_{\gamma 0}} \right) ^{\frac{1}{3\gamma
-2}}\sinh^{\frac{2}{3\gamma -2}}\left[ (3\gamma -2)\sqrt{\pi}
\sqrt{\frac{1- \Omega_{\gamma 0}}{\Omega_{\phi_0}}}
\,\frac{\phi}{m_{pl}} \right].
\end{equation}
On the other hand, inserting (\ref{pq17}) into (\ref{pq9}) we get
the expressions for the energy densities of the two components in
terms of $\phi$, that is
\begin{equation}
\label{dM} \rho(\phi) \ = \ \rho_{\gamma 0}\,\left(\frac{1-
\Omega_{\gamma 0}} {\Omega_{\gamma
0}}\right)^{\frac{3\gamma}{3\gamma -2}}\,
\sinh^{-\frac{6\gamma}{3\gamma - 2}}\left[ (3\gamma -
2)\sqrt{\pi}\sqrt{\frac{1- \Omega_{\gamma 0}}{\Omega_{\phi_0}}}\,
\frac{\phi}{m_{pl}}\right],
\end{equation}
\begin{equation}
\label{df1}\rho_{\phi}(\phi) \ = \ \rho_{\phi_0}\, \left( \frac{1-
\Omega_{\gamma 0}} {\Omega_{\gamma 0}}\right)^{\frac{2}{3\gamma
-2}}\, \sinh^{-\frac{4}{3\gamma -2}}\left[ (3\gamma -2)\sqrt{\pi}
\sqrt{\frac{1- \Omega_{\gamma 0}}{\Omega_{\phi_0}}}
\,\frac{\phi}{m_{pl}} \right].
\end{equation}

Now, substituting (\ref{df1}) into (\ref{V0}), one obtains the
potential $V(\phi)$, i.e.,

\begin{equation}
\label{pq19} V(\phi) \ = \ \frac{2}{3}\rho_{\phi_0}\, \left(
\frac{1- \Omega_{\gamma 0}} {\Omega_{\gamma
0}}\right)^{\frac{2}{3\gamma -2}}\, \sinh^{-\frac{4}{3\gamma
-2}}\left[ (3\gamma -2)\sqrt{\pi} \sqrt{\frac{1- \Omega_{\gamma
0}}{\Omega_{\phi_0}}} \,\frac{\phi}{m_{pl}} \right].
\end{equation}

Naturally, the behavior for different epochs may be obtained by an
appropriate choice of $\gamma$. In particular, taking $\gamma =1$ in
(\ref{pq19}), one may see that the potential $V(\phi)$ reduces to
the one found by Di Pietro and Demaret \cite{R9a}.

\section{Transition epoch}

Let us now consider the transition redshift, $z_{\star}$, at which
the Universe switches from deceleration to acceleration or,
equivalently, the redshift at which the deceleration parameter
vanishes. In order to derive the general expression of the
transition redshift let us consider the deceleration parameter
written as $q(R)=-R\ddot{R}/R$. Now, by considering the general
differential equation governing the behavior of the scale factor
$R(t)$ in the presence of a perfect fluid $\gamma$ plus the
``X-matter" energy, i.e., equation (\ref{geneq}), it is
straightforward to show that, the transition redshift is given by
\begin{equation}
z_{\star}=\left(\left[\frac{3}{2\Delta}(\gamma-\omega-1)-1\right]
\left(\frac{\Omega_{\phi_{0}}}{\Omega_{\gamma_{0}}}\right)
\right)^{-\frac{1}{3(1-\gamma+\omega)}}-1.
\end{equation}
As one may check, by assuming the values $\gamma=1$, $\omega=-1$,
$\Omega_{\phi_{0}}=0.7$ and  $\Omega_{\gamma_{0}}=0.3$ we obtain
$z_{\star}=0.66$ which is in fully agreement with the result for the
cosmic concordance $\Lambda$CDM model. In addition, by slightly
modifying the pressure term $\gamma=1+\epsilon$, where
$\epsilon<<1$, we have
\begin{equation}
z_{\star}=\left[(1-3\epsilon)\left(\frac{\Omega_{\phi_{0}}}{\Omega_{\gamma_{0}}}
\right)\right]^{\frac{1}{3(1-\epsilon)}}-1.
\end{equation}
In particular, if $\epsilon\sim0.1$ we find $z_{\star}\sim0.59$, in
accordance with a recent kinematic determination based on two
different Supernovae type Ia samples \cite{Riess,CL08}.

\section{Conclusions}

We have discussed FRW cosmologies with a perfect simple fluid plus a
dark energy component. If the Quintessence fluid is represented by a
scalar field with constant equation of state parameter, $\omega$,
the EFE determine univocally the form of the scalar field potential.
In other words, we cannot postulate simultaneously an arbitrary form
for the potential and the ``X-matter" condition.

The general solution of $V(\phi)$ has been explicitly derived by
assuming a FRW flat Universes for generic values of the pair of
parameters ($\gamma, \omega$). In this case, our general solutions
(Eqs. (\ref{pq25}) - (\ref{pq29})) tell us how the potential behaves
at different epochs. As remarked there, the asymptotic behavior of
the potential at early universe (radiation era) was not previously
derived by another authors. The explicit analytic solution for the
potential is accompanied by the others relevant quantities like
$R(\phi)$, or equivalently $\phi(R)$, $\dot \phi(R)$ and
$\rho_{\gamma}$ and $\rho_{\phi}$. Naturally, the cosmological model
discussed here may be useful for universes filled with only two
dominant components. The efficiency of the method has also been
exemplified by considering closed and hyperbolic FRW spacetimes for
particular values of the EOS parameter $\omega$ and arbitrary values
of $\gamma$.

In the search for a more realistic description  of the Quintessence
field presumably filling the Universe, models with more than one
type of fluid component, such as a combination of radiation and
non-relativistic matter must also be considered. The possibility of
a time-dependent $\omega$ and interacting scalar fields as discussed
by many authors in the literature \cite{vary,Hov07,Interac} need
also to be investigated in light of the method proposed in the
present work.

\vspace*{0.5cm}

\noindent{\large \bf Acknowledgments}

This work was partially supported by the Fondo de Desarrollo
Universitario, UNAC (Per\'u) and by the Conselho Nacional de
Desenvolvimento Cient\'{\i}fico e Tecnol\'ogico-CNPq (Brazilian
Research Agency). JASL is also partially supported by FAPESP (No.
04/13668-0).

\end{document}